\documentclass[12pt]{article}

\usepackage{graphicx,  amssymb, array, amsmath, float}

\usepackage{color}
\usepackage[authoryear]{natbib}
\usepackage{bm}
\usepackage{graphicx}
\usepackage{float}
\usepackage{subcaption}
\usepackage{multicol}
\usepackage{xr}
\usepackage{mathtools}
\usepackage{comment}
\usepackage{nameref}%try loading last
\usepackage{hyperref}
\usepackage{fancyvrb}
\usepackage{multicol}

\newcommand{\blind}{0}

\begin{document}

\def\spacingset#1{\renewcommand{\baselinestretch}%
{#1}\small\normalsize} \spacingset{1}

\if0\blind
{
\title{Bayesian Computing in the Undergraduate Statistics Curriculum}
\author{Jim Albert* and Jingchen Hu**\\
*Department of Mathematics and Statistics, Bowling Green State University\\
**Department of Mathematics and Statistics, Vassar College}

\maketitle
} \fi

\if1\blind
{
  \bigskip
  \bigskip
  \bigskip
  \begin{center}
    {\LARGE\bf Bayesian Computing in the Statistics and Data Science Curriculum}
\end{center}
  \medskip
} \fi

\bigskip

\begin{abstract}

Bayesian statistics has gained great momentum since the computational developments of the 1990s. Gradually, advances in Bayesian methodology and software have made Bayesian techniques much more accessible to applied statisticians and, in turn, have potentially transformed Bayesian education at the undergraduate level. This article provides an overview on the various options for implementing Bayesian computational methods motivated to achieve particular learning outcomes. For each computational method, we propose activities and  exercises, and discuss each method's pedagogical advantages and disadvantages based on our experience in the classroom.  The goal is to present guidance on the choice of computation for the instructors who are introducing Bayesian methods in their undergraduate statistics curriculum.

%We provide computing recommendations for different models of undergraduate Bayesian modules or courses, from introductory applied ones to advanced ones with more emphasis on theory.
%We conclude with computing recommendations for different models of the modern Bayesian classroom, from introductory applied courses to advanced courses with more emphasis on theory.

\end{abstract}

\noindent%
\vfill

\newpage
\spacingset{1.45} % DON'T change the spacing!

\section{Introduction}
\label{intro}
\subsection{Introducing Inference Using the Bayesian Paradigm}

Statistical inference is generally taught from a frequentist perspective. Students are introduced to inferential methods such as confidence intervals and tests of hypotheses, and all procedures are evaluated by their performance under repeated sampling. Bayesian thinking provides an alternative way of introducing statistical inference: We express our beliefs about the location of one or more parameters by a prior distribution and use Bayes' rule to update our beliefs about these parameters after observing data.

There are a number of attractive aspects of the Bayesian paradigm that motivate the teaching of Bayesian methods at the undergraduate level.  \cite{berry1997teaching} argues that science is essentially subjective and the Bayesian approach embraces this idea by allowing for the inclusion of subjective opinion in the construction of priors. Whether it is a simple one-parameter model or a complex multilevel model with many parameters, we follow Bayes' rule to update our prior beliefs after observing data, and all types of inferences are found by summarizing the posterior distribution. Moreover, since Bayesian inferences are made conditional on the observed data, Bayesian statements of confidence are more intuitive than the analogous  frequentist statements. For example, a Bayesian 90\% interval estimate is a fixed interval that covers the parameter with probability 0.90, and one can make a decision about a hypothesis such as $H: \theta \le \theta_0$ by computing the posterior probability that $H$ is true. In addition, prediction can play an important role in Bayesian analyses, as  there is a straightforward way of predicting future observations by predictive distributions.

\subsection{The Bayesian Computation Challenge}

In a Bayesian model, we collect an observation vector $\bm{y}$ distributed according to a sampling density $f(\bm{y} \mid {\bm \theta})$ depending on one or more parameters denoted by {\bf ${\bm \theta}$}, and our prior beliefs about $\bm \theta$ are stated in terms of a prior density $g(\bm \theta)$.  Once $\bm{y}$ is observed, the likelihood $L({\bm \theta})$ is the density $f(\bm{y} \mid {\bm \theta})$ viewed as a function of ${\bm \theta}$. All inferences about the parameters are based on the posterior density $g(\bm \theta \mid \bm{y})$, which is proportional to the product of the likelihood and the prior:
$$
g({\bm \theta} \mid \bm{y}) \propto L({\bm \theta}) g({\bm \theta}).
$$
In addition, we are typically interested in predictions of future observations $\tilde{\bm{y}}$.  We learn about the location of future data by means of the predictive density $p(\tilde{\bm{y}} \mid \bm{y})$ obtained by integrating the sampling density $p(\tilde{\bm{y}} \mid {\bm \theta})$ over the posterior density $g({\bm \theta} \mid \bm{y})$:
$$
p(\tilde{\bm{y}} \mid \bm{y})  = \int p(\tilde{\bm{y}} \mid {\bm \theta}) g({\bm \theta} \mid \bm{y}) d{\bm \theta}.
$$

For many Bayesian models, posterior and predictive distributions are not analytically available.  Therefore, one impediment  in teaching the Bayesian paradigm is the computational burden of posterior and predictive calculations.    %The computational methods are illustrated using several examples.% typical of what is presented in our teaching.

\subsection{Learning Outcomes at the Undergraduate Level}
\label{intro:LOs}
If an educator decides to introduce Bayesian inference at the undergraduate level, several general learning outcomes listed below could inform the teaching, computation, and assessment of a Bayesian module or a Bayesian course.

\noindent {\bf Prior construction.}  Students generally are not familiar with specifying probabilities that reflect their subjective beliefs about unknown quantities.  They will have experience in constructing prior distributions that reflect their beliefs about the locations of the parameters.

\noindent {\bf Implementation of Bayesian inference.}  Given a Bayesian model consisting of a sampling distribution and a prior, students will understand how to compute the posterior distribution and how to find marginal posterior distributions of parameters of interest to perform inference.

\noindent {\bf Process of learning from data.} Students will understand how the posterior distribution combines two sources of information: the prior distribution and the data.  In many situations, the posterior will be a compromise between the prior and the data where the weights depend on the relative information contained in the two sources.

\noindent {\bf Applications of the predictive distribution}. Students will understand various uses of predictive distributions. The prior predictive distribution is helpful in construction of a prior, while the posterior predictive distribution is helpful in predicting future data and judging the suitability of the Bayesian model.

\noindent {\bf Simulation-based inference and prediction.} Inferences in a Bayesian approach are various summaries of the posterior distribution.  Students will understand that these inferential summaries can be approximated by simulated random samples from the posterior probability distribution.  In addition, they will understand how simulated samples can be drawn from the predictive distribution.

\noindent {\bf Bayes in popular statistical methods}. Students will understand how Bayesian models are implemented and how posterior and predictive distributions are computed in popular statistical methods, such as regression and multilevel modeling.

In this paper, we provide an overview of  available  computational strategies for an undergraduate Bayesian module or course, guided by this set of learning outcomes.  A number of papers such as \citet{casella1992explaining} and \citet{chib1995understanding} have focused on tutorials on specific Bayesian computational algorithms.  This paper has a different focus as it compares and contrasts different computational algorithms towards the goal of communicating the learning outcomes.

\subsection{A Selective History of Bayesian Computation}

In the 1960's there was an active interest in the practice of Bayesian learning methods.  \citet{raiffa1961applied}
was one of the early texts to describe the use of conjugate priors for exponential family distributions such as the normal, binomial, exponential, and Poisson.  
Other books providing descriptions of conjugate priors include \citet{winkler1972introduction},  \citet{lee1997bayesian}, and \citet{martz1982bayesian}.

In the 1960's, due to the conceptual simplicity of Bayesian thinking, there were efforts to introduce Bayesian inference at a non-calculus level.  One attractive way of introducing Bayes was to use discrete priors and use Bayes' rule to update prior opinion.
\citet{blackwell}, \citet{schmitt}, \citet{phillips1973bayesian}, and \citet{berry1996} are examples of introductory statistics textbooks that present inference for standard sampling distributions from a Bayesian viewpoint using discrete priors.

There were several important developments in Bayesian computation in the 1980's.  \citet{smith1985implementation} describe a general method for approximating Bayesian integrals using adaptive quadrature methods.  \citet{tierney1986accurate} describe accurate methods for approximating summaries of posterior distributions using Laplace expansions.

Statisticians became aware of Markov chain Monte Carlo (MCMC) methods through the seminal paper \citet{GelfandSmith1990JASA} that introduced Gibbs sampling for simulating from posterior distributions.  At the same time, \citet{gelfand1990illustration} illustrate the application of Gibbs sampling for Bayesian fitting for a range of normal sampling models.  Elementary expositions of Gibbs sampling and the general Metropolis-Hastings algorithm are found in  \citet{casella1992explaining} and \citet{chib1995understanding}.

\subsection{Two Examples}
\label{intro:examples}

The general goals of this paper are to provide a broad perspective of the computational methods currently available in a Bayesian analysis and present guidance for the choice of methods for introducing Bayesian thinking at the undergraduate level. We will present each computational method with a discussion of pros and cons from the pedagogical perspective. Moreover, we will illustrate most of the methods within the context of the following two examples.
%The computational methods will be illustrated within the context of two examples.

\subsubsection*{Visits to an Emergency Department}

Suppose a hospital administrator is studying the pattern of frequency of arrivals to the emergency department (ED) of her hospital.  She focuses on the variable $Y$,  the count of new arrivals to the ED during the evening hour between 9 pm and 10 pm.  She assumes that the number of arrivals, $Y_1, ..., Y_n$, for $n$ distinct one-hour periods between 9 pm and 10 pm are independent Poisson($\lambda$) random variables where $\lambda$ represents the average number of arrivals in this one-hour period.  The likelihood function is given by
\begin{equation}\label{eq1}
L(\lambda) \propto \prod_{i=1}^n \exp(-\lambda) \lambda^{y_i} = \exp(-n \lambda) \lambda^{\sum y_i}, \, \, \lambda > 0.
\end{equation}

Since the administrator has previously worked at hospitals of similar size, she has some prior knowledge about $\lambda$ before any data is collected and wishes to construct a prior density $g(\lambda)$ that reflects this information.  After observing data for 10 periods, she is interested in using the posterior distribution to construct an interval estimate for $\lambda$.  Also she would like to use the posterior predictive distribution to predict the total number of ED arrivals for 10 future days in the particular period between 9 pm and 10 pm, so that data-driven decisions can be made about staffing.

\subsubsection*{Comparing Proportions of Facebook Users Using a Logistic Model}

 Suppose a survey is given to a sample of male and female college students regarding their use of Facebook.  Of $n_M$ men sampled, $y_M$ are frequent users of Facebook, and $y_W$ out of $n_W$ women sampled are frequent Facebook users.  Let $p_M$ and $p_W$ denote respectively the proportions of college men and college women who are frequent users of Facebook.  One can relate the proportions to the gender variable by means of the following logistic model.
 \begin{equation} \label{eq2}
\begin{split}
\log \left(\frac{p_M}{1-p_M} \right) & =  \beta_0 \\
\log \left(\frac{p_W}{1-p_W} \right) & = \beta_0 + \beta_1
\end{split}
\end{equation}

In the model in Equation (\ref{eq2}), the slope parameter $\beta_1$ represents the log odds ratio that measures the difference between the two proportions.  Suppose that we believe \emph{a priori} that the proportions of college men and college women who use Facebook frequently are similar in size. Specifically, we believe that $\beta_1$ falls in the interval $(-0.5, 0.5)$ with probability 0.5, but we do not have much knowledge about values of $\beta_1$ outside of this interval.  This prior information can be matched by
assigning $\beta_1$ a Cauchy density with location 0 and scale 0.5\footnote{Note that our prior belief has two components. The first prior belief is that the probability that $\beta_1$ is within 0.5 is equal to 0.5, and the second prior belief is that little is known about the location of $\beta_1$ outside of the interval (-0.5, 0.5).  A Cauchy density is a better match to this prior information than the normal density since the Cauchy has flatter tails than the normal reflecting lack of knowledge of prior information in the tails.}. The intercept parameter $\beta_0$ is assigned a normal prior with mean 0 and standard deviation 100 that reflects little knowledge about the location of this parameter.  The posterior density of the coefficient vector $\bm \beta = (\beta_0, \beta_1)$ is proportional to

\begin{equation}\label{eq3}
g(\bm \beta \mid data) \propto \frac{\exp(\beta_0 y_M)}{[1 + \exp(\beta_0)]^{n_M}}
\frac{\exp((\beta_0 + \beta_1) y_W)}{[1 + \exp(\beta_0 + \beta_1)]^{n_W}} 
\, \, \frac{1}{0.25 + \beta_1^2} \exp\left(-\frac{\beta_{0}^2 }{2 (100)^2}\right).
\end{equation}

\smallskip

\noindent Note that the posterior density in Equation (\ref{eq3}) is not a familiar functional form, so some type of numerical method is required to summarize the posterior.

%\subsection{Plan of the Paper}

%The general goal of this paper is to provide a broad perspective of the computational methods currently available in a Bayesian analysis and present guidance for the choice of method for introducing Bayesian thinking at both undergraduate and graduate levels.  

The reminder of this article is organized as follows. Section \ref{nonsims} summarizes and discusses various non-simulation approaches to Bayesian computing, including the use of discrete prior distribution, conjugate priors, and normal approximations.  Section \ref{sims} introduces and discusses simulation-based Bayesian computations. Section \ref{MCMC} is devoted to popular MCMC samplers, and focuses on the uses of Gibbs sampler, Metropolis algorithm, and Just Another Gibbs Sampler (JAGS) in teaching. For each introduced method, we propose activities and / or exercises, and discuss the method's advantages and disadvantages. The article ends with Section \ref{conclusion} with several concluding remarks.

% including the Gibbs sampler, the Metropolis-Hasting algorithm, and the Hamiltonian Monte Carlo sampler.  We believe that the Gibbs sampler and the Metropolis algorithm are attractive for introducing MCMC methods and we describe the advantages and disadvantages of each method from a pedagogical perspective.  This section also reviews methods for coding MCMC samplers such as writing R functions, writing a model script to be used by another program, or using a wrapper function that implements MCMC fitting for specific Bayesian models.  Section 5 concludes the paper with some summary remarks on effective Bayesian computational methods for students with different backgrounds.

\section{Non-Simulation Approaches}
\label{nonsims}

\subsection{Discrete Bayes}
\label{nonsims:discrete}

Bayesian thinking  can be introduced by means of discrete distributions.  Define a {\it model} to be a particular characteristic of a population.  A model can be a parameter such as a population mean or proportion, or other population measures.  
Suppose we can construct a list of model values \{$M_j$\} with associated prior probabilities \{$P(M_j)$\}.  Let $\bm{y}$ represent a vector of observations that can shed some light on the models.  Then by Bayes' rule,  the posterior probability of model $M_j$ is given by
$$
P(M_j \mid \bm{y}) \propto P(M_j) L(M_j),
$$
where $L(M_j)$, the likelihood,  is the probability of the observed data $\bm{y}$ given the model value $M_j$.  

The discrete Bayes approach is typically illustrated in undergraduate statistics courses with the familiar scenario of disease testing. Suppose a person is concerned that she has a rare disease and takes a blood test that will help determine if she has the disease.  Unfortunately, the blood test is not completely reliable -- the error rates of false positives and false negatives  are provided.  In this setting, we apply Bayes' rule to determine the probability she has the disease given a positive blood test result, where the prior distribution usually comes from the percentage of the rare disease in the population. 

To provide an introduction to the specification of a prior distribution, we illustrate the discrete Bayes approach applied to our example of visits to the ED.  The administrator specifies a list of plausible values, $\lambda_1$, ..., $\lambda_k$, of the mean arrival rate $\lambda$ and associated weights, $w_1, ..., w_k$, that reflect her belief in the relative likelihoods of these values.  A weight of say, 10, can be assigned to the most likely value of $\lambda$, a weight of 5 can be assigned to another $\lambda$ value that is half as likely as the most probable value, and so on.  Once all of these weights are assigned, they are easily converted to prior probabilities by dividing each weight by the sum of the weights.

%The discrete Bayes computations can be illustrated by a Bayes' table displayed in Table \ref{tab:discrete}.  
Table \ref{tab:discrete} is a Bayes' table that can be used to illustrate the discrete Bayes computations. 
The plausible values of the Poisson mean, $\{\lambda_1, \cdots, \lambda_k\}$, and the corresponding prior probabilities, $\{p_1, \cdots, p_k\}$, are listed respectively in the  ``Model" and ``Prior" columns.   The data consists of the number of ED arrivals in $n$ periods and the ``Likelihood" column gives the likelihood  for each of the $\lambda_j$  ($j = 1, \dots, k$) values.  We compute posterior probabilities in two steps.  First, we compute the product of the prior probability and the likelihood for each value of $\lambda$, presented in the ``Product" column. Second, we normalize these products (by dividing each product by the sum of the products) to obtain the posterior probabilities, presented in the ``Posterior" column.

\begin{table}[ht]
\centering
\caption{Bayes table representation for learning about a discrete-valued parameter.}
\label{tab:discrete}
\begin{tabular}{ccccc}
  \hline
Model & Prior & Likelihood & Product & Posterior \\ 
  \hline
 $\lambda_1$   & $p_1$ & $L(\lambda_1)$  & $p_1 L(\lambda_1)$ & $p_1 L(\lambda_1) / S$\\ 
 $\lambda_2$ & $p_2$ & $L(\lambda_2)$  & $p_2 L(\lambda_2)$ & $p_2 L(\lambda_2) / S$ \\ 
 ... &...  &... & ... & ... \\
 $\lambda_k$ & $p_l$ & $L(\lambda_k)$  & $p_k L(\lambda_k)$ & $p_k L(\lambda_k) / S$ \\ 
   \hline
Sum &  & & $S$ &  \\
\end{tabular}
\end{table}

In our example, suppose the hospital administrator believes $\lambda$ values of \{3, 3.5, 4, 4.5, 5\} are possible  by looking at evening ED visit summaries for hospitals serving similar-size communities.  In addition, these summaries make her believe that the rate value of $\lambda = 4$ is most likely and values of $\lambda = 3.5$ and 4.5 are half as likely as the most likely value.
Based on these thoughts, she assigns these values the respective prior probabilities \{0.1, 0.2, 0.4, 0.2, 0.1\}.  The number of ED visits $\bm{y} = (y_1, ..., y_n)$ for $n = 10$ periods is observed, and one computes $\sum y_i = 31$.  Table \ref{tab:discreteexample} presents the Bayes' table calculations.  Note how the prior opinion about the average number of ED visits has been adjusted in light of the data.  Initially the administrator thought $\lambda$ was likely between 3.5 and 4.5.  After observing the data, she believes the most likely value of $\lambda$ is 3.5. Moreover, she can construct a probability interval for $\lambda$ by placing individual values of $\lambda$ into a ``probability bin", starting with the most likely value and stopping when the total probability in the bin exceeds 0.95.  Using this method, a 95\% posterior probability interval that $\lambda$ lies is [3.0, 4.0], by calculating $0.241 + 0.386 + 0.327 = 0.954$.

\begin{table}[ht]
\centering
\caption{Bayes table representation for ED example.}
\label{tab:discreteexample}
\begin{tabular}{rrrrr}
  \hline
 $\lambda$ & Prior & Likelihood & Product & Posterior \\ 
  \hline
3.0 & 0.1 & 57.8 & 5.78 & 0.241 \\ 
  3.5 & 0.2 & 46.3 & 9.26 & 0.386 \\ 
4.0 & 0.4 & 19.6 & 7.84 & 0.327 \\ 
  4.5 & 0.2 & 5.1 & 1.02 & 0.042 \\ 
5 & 0.1& 0.9& 0.09 & 0.004 \\ 
   \hline
\end{tabular}
\end{table}

 \noindent {\underline{\bf Activity 1 - Learning About a Proportion}}: The process of constructing a prior distribution can be challenging for students, especially since they have had little experience specifying subjective probabilities.  The Discrete Bayes Activity: ``Learning About a Proportion" in Section 1.1 of the Supplementary Material is designed to help students specify a discrete prior for a proportion of interest and make inferences about the proportion from results from a sample survey.  The posterior calculations are facilitated by a Javascript app.

There are a number of attractive features of the discrete Bayes approach from a pedagogical  perspective.  First, the discrete Bayes approach provides a simple introduction to the construction of a prior.  Students typically have little experience placing probabilities on unobserved quantities such as a population mean, while specifying probabilities on a small set of plausible set of parameters is a task that is possible for many students.  (\cite{albert1998using} describes students' construction of discrete priors in a project setting.) Second, we have a clear algorithm for computing  posterior probabilities by taking products of prior probabilities and likelihood values.  
By comparing prior and posterior distributions by the inspection of modes or means, we can see how our prior opinion has been modified with the information in the data.  Moreover, posterior inference is performed by summarizing a discrete probability distribution.  As illustrated in the ED example, a  Bayesian interval estimate is found by collecting the values of $\lambda$ with the largest posterior probabilities.  Last, predictive probabilities are expressible as sums.  For example, if the administrator wishes to predict the number of ED visits $y^*$ in a future period, the posterior predictive mass function is given by
$$
f(y^*) = \sum_{j=1}^k p(\lambda_j) f(y^* \mid \lambda_j),
$$
where $\{p(\lambda_j)\}$ are the current (posterior) probabilities and $f(y^* \mid \lambda_j)$ is the Poisson sampling density given the mean $\lambda_j$.  Suppose the administrator wishes to predict the probability that there are no visits to the ER in a future one-hour period between 9 and 10 pm.  Using the posterior probabilities from Table \ref{tab:discreteexample}, the predictive probability of $y^* = 0$ is given by
\begin{eqnarray}
f(0)&  = & 0.241 \exp(-3.0) + 0.386 \exp(-3.5) + 0.327 \exp(-4.0) + 0.042 \exp(-4.5) \nonumber \\
 & + &  0.004 \exp(-5)  = 0.030.
\end{eqnarray}
The administrator concludes that it is rare to observe no ER visits during this evening hour.

%\subsubsection*{Discrete approximation}

%One attractive way of learning about a continuous-valued parameter ${\bm \theta}$ from a Bayesian viewpoint is through a discrete Bayes approximation.  
%For a given sampling model $f(y \mid {\bm \theta})$  and prior density $g({\bm \theta})$, it may be difficult to compute summaries of the posterior density analytically. However, one has the approximation
%$$
%g({\bm \theta} \mid y) \propto g({\bm \theta}) f(y \mid {\bm \theta})  \approx \sum_{j=1}^N g({\bm \theta}_j) f(y \mid {\bm \theta}_j),
%$$
%where ${\bm \theta}_1, ..., {\bm \theta}_N$ is a fine grid of $N$ values of ${\bm \theta}$ that covers the range of values where the posterior density has most of its probability content.

%From a computational perspective, this discrete approach is very appealing.  In R, for example, one can implement Bayesian model by use of three vectors, one containing the parameter values, one containing the prior values, and a third computing the likelihoods (facilitated by the availability of functions like \texttt{dnorm()} and \texttt{dbeta()} for common distributions).  

Despite its computational simplicity, there are challenges in using the discrete Bayes approach.  Conceptually the discrete Bayes approach can be used for posteriors of multiple parameters where each parameter is assigned values on a grid.  However, the number of  posterior evaluations grows exponentially as a function of the number of parameters, therefore the use of the discrete Bayes approach may be limited to a small number of parameters.

\subsection{Conjugate Analyses}
\label{nonsims:conjugate}

The use of discrete Bayes in the  ED visits example has another clear limitation: we are using a discrete distribution to approximate beliefs about a continuous-valued parameter $\lambda$.  For  
 a one-parameter exponential family distribution such as the Poisson, there exists an attractive conjugate Bayesian analysis where the prior and posterior densities have the same functional form.  
 
In our ED visits example, if a random sample $\bm{y} = (y_1, ..., y_n)$ is taken from a Poisson distribution with mean $\lambda$, then a gamma prior is conjugate. That is, if the mean number of visits $\lambda$ is assigned a gamma prior with shape $\alpha$ and rate $\beta$ proportional to
$$
g(\lambda) \propto \lambda^{\alpha - 1} \exp(-\beta \lambda),
$$
 the posterior density will also of the gamma functional form with updated parameters $\alpha_1 = \alpha + \sum_{j=1}^n y_j$ and $\beta_1 = \beta + n$.

There are several advantages to using a conjugate prior in teaching Bayesian methods.

\noindent {\bf Ease of specifying prior densities}.  The use of a conjugate prior simplifies the process of choosing a prior.   In our ED visits example, by using a gamma prior, the hospital administrator only needs to specify the gamma parameters, $\alpha$ and $\beta$, that reflect the location and spread of the distribution.  One convenient way to specify these parameters is to first specify a value of the prior mean $\mu = \alpha / \beta$ and  then specify the parameter $\beta$ representing the strength of this belief in the  prior mean expressed in terms of the size of a ``prior" sample.  For example, if the administrator's ``best guess" at $\lambda$ is 4, and she believes this guess is worth about 20 observations, then she could use a gamma prior with parameters $\alpha = \mu \beta = 80$ and $\beta = 20$.  

\noindent {\bf Ease of computing posterior summaries.}  With the use of a conjugate prior, the posterior means and posterior standard deviations have closed form expressions.   These simple expressions help in communicating how the prior information and data are combined in the posterior distribution.  For example, the posterior mean of $\lambda$ for our example has the form
$$
E(\lambda \mid y_1, \cdots, y_n) = \frac{\sum_{i=1}^n y_i + \alpha}{n +\beta} = \left( \frac{n}{n+\beta}\right) \bar y + \left(\frac{\beta}{n + \beta}\right)\mu,
$$
which is a weighted average of the sample mean $\bar y$ and the prior mean $\mu$. Moreover, we can analytically calculate that the weights, $\frac{n}{n + \beta}$ and $\frac{\beta}{n + \beta}$, are determined by the relationship between the prior sample size $\beta$ and data sample size $n$.

\noindent {\bf Straight-forward inference.} Exact summaries of the posterior density are available since the posterior has a familiar functional form.  In our ED visits example, posterior probabilities can be found using the R function  \texttt{pgamma()} and probability intervals can be found using gamma quantiles found using the R function \texttt{qgamma()}.  In our example, $n=10$ periods are observed and the sum $\sum y_i = 31$, so the posterior of $\lambda$ is Gamma($80 + 31, 20 + 10$).  We can construct a 90\% interval estimate for $\lambda$ by extracting the 5th and 95th percentiles from the Gamma(111, 30) distribution:
\begin{verbatim}
qgamma(c(0.05, 0.95), shape = 111, rate = 30)
[1] 3.142 4.296
\end{verbatim}

\noindent {\bf Closed-form predictive densities.} Also due to the conjugate structure, exact analytical expressions exist for the predictive density.  This  facilitates the construction of prediction intervals for future data.% {\color{blue} MH: do we want to do a demo for prediction?}

\noindent {\underline{\bf Activity 2 - Did Shakespeare Use Long Words?}}: The activity ``Did Shakespeare Use Long Words?" in Section 1.2 of the Supplementary Material illustrates using a conjugate prior to learn about the proportion of long words used by William Shakespeare in his plays.  Students specify the median and 90th percentile of their prior and a Shiny app is used to find the shape parameters of the matching beta prior.  Students collect data from one of Shakespeare's plays and performs inference from the beta posterior distribution.

The main challenge of teaching conjugate analyses is the necessity of correctly deriving the posterior distribution, which requires a solid background of calculus and probability. We recommend starting with simple one-parameter models, such as the gamma-Poisson, beta-binomial and normal-normal models.  A possible computing challenge is that some students may need to learn R functions such as \texttt{qgamma()}  to perform relevant posterior inferences. On this topic, we recommend learning by examples. For example, once the gamma-Poisson conjugate analysis is introduced, the students can be given a lab on the beta-binomial conjugate analysis, with sample scripts to practice using appropriate R functions for posterior inferences.

\subsection{Normal Approximation}
\label{nonsims:normal}

An alternative non-simulation computational method is based on approximating the posterior with a normal curve. (See \citet{tierney1986accurate}  for a discussion of related approximations.)  An algorithm such as Newton's method can be used to find the posterior mode $\tilde {\bm \theta}$, the value where the posterior density achieves its maximum value.  Then we obtain the normal approximation
$$
g({\bm \theta} \mid \bm{y}) \approx \textrm{Normal}(\tilde {\bm \theta}, \bm V),
$$
where the variance-covariance $\bm V$ is estimated by the behavior of the posterior curve about the modal value. 

%From a computational perspective, this is an attractive approximation due to the nice properties of the multivariate normal distribution.  For example, the marginal distributions of the posterior are approximated by normal curves, and posterior intervals are easily computed using normal quantiles. 

\begin{figure}[H]
  \centering
\includegraphics[scale=0.5]{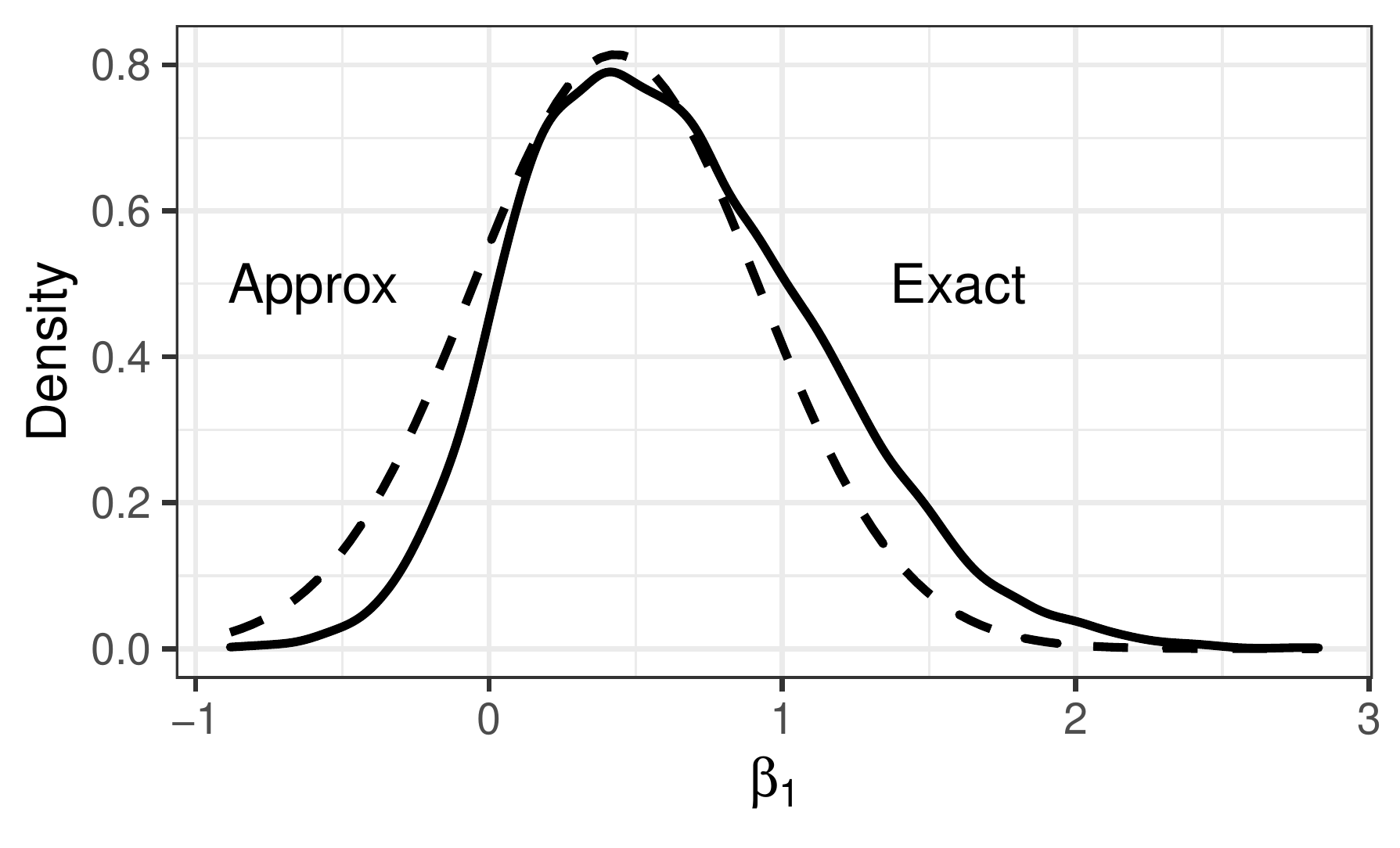}
\caption{\label{fig:exactapprox} Comparison of ``exact" grid and normal approximation to the marginal posterior of the slope parameter $\beta_1$ in the two-group logistic model example.}
\label{fig:normal}
\end{figure}

We illustrate the use of the normal approximation for the two-group logistic model example introduced in Section \ref{intro:examples}. The R code for this example is provided in Section 2 in the Supplementary Material.
Suppose we observe $y_M = 8$ Facebook users in a sample of $n_M = 30$ men, and $y_W = 15$ Facebook users in a sample of $n_W = 30$ women. We find  that $\bm \beta = (\beta_0, \beta_1)$  has the following bivariate normal approximation:
\begin{gather}\label{eq:normalapprox}
 \begin{bmatrix} \beta_0 \\ \beta_1 \end{bmatrix}
 \approx
 \textrm{Normal}\left(
 \begin{bmatrix}
 $-0.696$ \\ 0.431
 \end{bmatrix},
  \begin{bmatrix}
   0.137 &
  $-0.126$ \\
   $-0.126$ &
   0.239 
   \end{bmatrix}\right).
\end{gather}
By inspecting Equation (\ref{eq:normalapprox}),  one sees that the log odds ratio $\beta_1$ is approximately normal with mean 0.431 and standard deviation $\sqrt{0.239} = 0.489.$ This  approximation is displayed as the ``Approx" curve in Figure \ref{fig:normal} and contrasted with an ``Exact" curve based on quadrature methods. We observe that this normal approximation does not match the right skewness of the exact posterior density.  However, it provides a reasonably accurate and easy to compute approximation to the posterior density of interest.

\noindent {\underline{\bf Activity 3 - Normal Approximation to Posterior}}: The activity ``Normal Approximation to Posterior" in Section 1.3 of the Supplementary Material explores the use of this normal approximation when students are interested in learning about the standard deviation of a distribution of computing times.

There are several advantages to the use of a normal approximation in teaching Bayesian methods.

\noindent {\bf A general approach.} The normal approximation is a general computing approach applicable to any Bayesian model with arbitrary specifications of the sampling density and prior.  In particular, we do not have to use a conjugate prior to represent prior opinion. The two-group logistic model is an example of a Bayesian posterior distribution which can not be represented by a standard parametric family.

\noindent {\bf Nice properties of multivariate normal distribution.}
Another advantage of the multivariate normal approximation is that it gives normal approximations to the marginal posterior densities. In our example, it was straightforward to obtain a normal approximation to the log odds ratio $\beta_1$ from the bivariate normal approximation of the regression vector.  Also, by use of properties of the multivariate normal, we can obtain approximations for marginal posterior distributions of linear functions of the parameters.

\noindent {\bf Can apply simulation methodology.}
Since it is convenient to simulate from a multivariate normal distribution, the approach is amenable to the simulation approach for Bayesian computation (to be described in Section \ref{sims}). To illustrate the use of simulation, suppose one is interested in estimating the proportion $p_W$ in our example that is given by
\begin{equation}
p_W = \frac{\exp(\beta_0 + \beta_1)}{1 + \exp(\beta_0 + \beta_1)}.
\label{eq:pW}
\end{equation}
One can obtain a simulated sample from the posterior of $p_W$ by simulating a large number of values from the normal approximation to $(\beta_0, \beta_1)$ in Equation (\ref{eq:normalapprox}) and then substituting these simulated draws in Equation (\ref{eq:pW}).
{\it Statistical Rethinking} \citep{Rethinking2016book} is an example of a modern applied Bayesian textbook that uses this normal approximation in introducing posterior simulation in a regression setting. 

One challenge in teaching the normal approximation is getting students to understand the limitations of the method.  For example, the normal approximation implicitly assumes that the parameter is real-valued which may not be suitable in cases where the parameter has bounded support such as a standard deviation or a proportion.  Through examples, students can be shown that the normal approximation may not be suitable for these problems where the parameter has bounded support.  To help understand that the normal curve is only an approximation, it is helpful to consider situations where one can compare different computational methods, say normal approximation and simulation, for evaluating a specific posterior summary such as a posterior mean.

\section{Simulation Approaches}
\label{sims}

\subsection{Overview}
\label{sims:overview}

Simulation provides a general strategy in Bayesian computation. Without loss of generality, consider a single-parameter model with parameter $\theta$ and a vector of observations $\bm y$. Suppose we can simulate a sample of $S$ values $\theta^{(1)}, ..., \theta^{(S)}$ from the posterior density $g(\theta \mid \bm{y})$.  Then we can compute various summaries of the posterior distribution by summarizing the corresponding simulated draws.  For example, the posterior mean of $\theta$, $E(\theta \mid \bm{y})$, is approximated by the sample mean of simulated values
$$
E(\theta \mid \bm{y}) \approx \frac{\sum_{s = 1}^S \theta^{(s)}}{S}.
$$
If we wish to construct,  say a 90\% credible interval for $\theta$, this interval is approximated by ($q_L, q_U$), where $q_L$and $q_U$ are respectively the 5th and 95th percentiles of the sample of simulated draws \{$\theta^{(s)}, s = 1, \cdots, S$\}.

Simulation can be applied for each of the computational approaches described in Section \ref{nonsims}.  If we approximate a continuous-valued posterior $g(\theta \mid \bm{y})$ by a discrete distribution, we can simulate from the posterior by taking a random sample with replacement from the values $\{\theta_j\}$ with probabilities proportion to $\{g(\theta_j)\}$.
Conjugate models allow for convenient simulation of posterior densities, as the posterior distribution is often a familiar functional form and algorithms are available for simulating from these distributions.  If a normal distribution is used to approximate the posterior, then simulated draws from the normal can be used to perform inference.  

From a pedagogical perspective, simulation approaches can be introduced to Bayesian methods with conjugate analyses, where students evaluate and compare exact solutions and simulated solutions, further instilling their understanding of conjugate analyses. Many inferential questions can be answered in a straightforward manner by summarizing simulated posterior draws and functions of posterior draws, which serves not only as a means to solve such inferential questions, but also as great exercises and practices for students to understand the key role that simulation approaches play in Bayesian methods. Moreover, it builds a foundation for students' learning of Markov chain Monte Carlo (MCMC) methods later in the course. We now turn to the ED visits example with simulation approaches to illustrate these advantages.

\subsection{Example: ED Visits}
\label{sims:gammaPoisson}

In the ED visits example with a conjugate analysis of Section \ref{nonsims:conjugate}, it is assumed that the visit counts $\bm{y} = (y_1, \cdots, y_n)$ were a random sample from a Poisson distribution where the mean $\lambda$ is assigned a Gamma$(\alpha, \beta)$ prior.  In the example, suppose $\alpha = 80, \beta = 20$ for the gamma prior, and $n = 10$ and $\sum y_i = 31$ from the data, and the posterior has a gamma distribution with parameters $(\alpha + \sum_{i=1}^{n}y_i, \beta + n) = (111, 30)$.

Suppose we wish to find a 90\% interval estimate for $\lambda$.  A simulation-based approach here is to simulate a large number of draws  from the posterior distribution and approximate a 90\% interval by finding the 5th and 95th percentiles of this simulated sample. These steps can be done using the R functions \texttt{rgamma()} and \texttt{quantile()}:
\begin{verbatim}
S <- 1000
lambda_draws <- rgamma(S, shape = 111, rate = 30)
quantile(lambda_draws, c(0.05, 0.95))
5%      95% 
3.137 4.248
\end{verbatim}
The obtained simulation-based 90\% interval of [3.137, 4.248] closely approximates the exact interval of [3.142, 4.296] obtained  using the \texttt{qgamma()} function in Section \ref{nonsims:conjugate}.

% Introducing the simulation-based methods for conjugate models such as gamma-Poisson, provides the space for experimentation and discussion of the effect of $m$ on Monte Carlo approximation accuracy. 

%In posterior analysis of conjugate model such as gamma-Poisson, it is not necessary to use simulation since exact gamma posterior summaries are available. However, from a pedagogical perspective, introducing the simulation-based methods provides a good way to introduce the use of simulation in a Bayesian analysis, which will become the main approach in advanced models where exact solutions are not available. Moreover, it is helpful to contrast simulation-based posterior summaries with the exact values, as what we have just done. Along these lines, accuracy of simulation-based approximation to exact solutions can be discussed, and the link between the approximation accuracy and the simulation sample size $m$ can be introduced. One can design a computing lab or a homework question to invite students to try out different values of $m$ and to comment on its impact on the approximation accuracy.
 
There are several attractive aspects of using simulation in teaching Bayesian methods.

\begin{figure}[H]
  \centering
\includegraphics[scale=0.3]{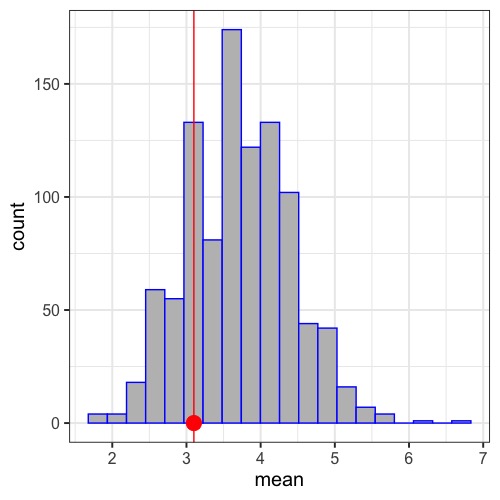}
\caption{\label{fig:ppchecks} Histogram of sample means of predicted samples, compared to the actual sample mean (vertical red line).}
\end{figure}

\noindent {\bf Learning about functions of parameters.} If we are interested in a function $h(\lambda)$, we can simulate from the marginal posterior distribution of $h(\lambda)$ by applying this function on the vector of simulated draws of $\lambda$. For example, suppose we are interested  in $Pr(y \leq 2 \mid \lambda)$, the probability (conditional on $\lambda$) that a particular one-hour period has at most 2 ED visits. This probability can be expressed as a function of $\lambda$:  $h(\lambda) = Pr(y \leq 2 \mid \lambda) = Pr(y = 0   \mid \lambda) + Pr(y = 1  \mid \lambda) + Pr(y = 2  \mid \lambda) = \exp(-\lambda)(1 + \lambda + \lambda^2/2)$. We can sample from the posterior of $h(\lambda)$ by applying this function on a large number of simulated posterior draws of $\lambda$.  We illustrate below this process and computing the posterior mean of $Pr(y \leq 2 \mid \lambda)$: 

\begin{verbatim}
S <- 1000
lambda_draws <- rgamma(S, shape = 111, rate = 30)
h_lambda <- exp(-lambda_draws) * (1 + lambda_draws + lambda_draws^2/2)
mean(h_lambda)
0.293
\end{verbatim}

\noindent {\bf Posterior predictive model checking.} A general way of checking the suitability of a Bayesian model is to explore if the observed data is consistent with replicated data simulated from the posterior predictive distribution.  This model checking approach is practical given the ease of simulating replicated datasets. We simulate a large number of predicted samples of the same size as the data sample, compute a key statistic in each predicted sample, and compare the collection of predicted statistics with the key statistic in the data sample. For example, if the statistic is the sample mean, we simulate 1000 samples of size 10 from the posterior predictive distribution and  compute the sample mean for each sample. (See the R code for implementing this simulation below.) Figure \ref{fig:ppchecks} displays a histogram of the simulated sample means and displays the actual sample mean of 3.1 in the ED visits dataset as a vertical line. The actual sample mean is in the bulk of the distribution of the simulated sample means, indicating that the chosen Bayesian model is suitable. 

%\begin{multicols}{2}

\begin{verbatim}
one_pp_sim <- function(n){
  lambda_draw <- rgamma(1, shape = 111, rate = 30)
  y_pred <- rpois(n, lambda_draw)
  mean(y_pred)
}
sample_means <- replicate(1000, one_pp_sim(10))
\end{verbatim}

\section{MCMC}
\label{MCMC}

\subsection{Popular MCMC Samplers}
\label{MCMC:intro}

In non-conjugate, multi-parameter  Bayesian models,  exact solutions to the posterior distribution are usually analytically unavailable.  Therefore, we need to rely on simulation-based computations for posterior estimation, and a popular class of computation techniques is called Markov chain Monte Carlo (MCMC).

In MCMC, we construct a specific Markov chain to step through a high-dimensional posterior probability distribution.  Informally, we are constructing a type of random walk that searches for locations where the posterior distribution has high probability content.  Under general conditions, the Markov chain will approach, as the number of steps gets large, an equilibrium distribution that is equivalent to the posterior distribution of interest. Popular MCMC samplers for Bayesian inference include the Gibbs sampler, the Metropolis-Hastings algorithm, and the Hamiltonian Monte Carlo algorithm (HMC).   

A Gibbs sampler iteratively samples one parameter at a time given its full conditional posterior distribution, defined as the parameter's distribution conditional on values of the remaining parameters.  

The family of Metropolis-Hastings algorithms provides a general way of implementing a Markov chain in situations where the full conditional posterior distributions are not recognizable.  In a Metropolis-Hastings algorithm, a proposal distribution is used to select candidate simulated draws, and we decide to move to the candidate draw or remain at the current simulated value depending on an acceptance probability.  The Metropolis algorithm is a special case of the Metropolis-Hastings algorithm that uses a symmetric proposal distribution. 

The HMC is a more efficient MCMC sampler that exploits information about the geometry of the typical set designed for  generating efficient draws of the posterior distribution  for sufficiently well-behaved target posterior distributions \citep{HMC2011, HMC2017Betancourt}.  A popular MCMC software program Stan \citep{carpenter2017stan} interfaces with many programming languages such as R.

An MCMC sampler will only converge to the target posterior distribution in theory and  the collected MCMC draws are an approximation to the unknown joint posterior distribution.  When implementing an MCMC sampler, we want to know how long the sampler needs to run until it reaches the space where the posterior has most of its probability.  In addition, we wonder about the number of iterations  needed to be collected to obtain accurate estimates at posterior summaries of interest, and this concern relates to the correlations of the successive MCMC sampled values.  The collection of diagnostic methods used to address the questions of MCMC convergence are called MCMC diagnostics \citep{mengersen1999mcmc}.

From a pedagogical perspective, we believe that students should be introduced to MCMC algorithms at an appropriate depth. Before using ``black box"  MCMC software for advanced multi-parameter Bayesian models that they encounter, we advocate first introducing MCMC algorithms for relatively simple Bayesian models using self-written MCMC samplers. For students with modest coding background, once full conditional posterior distributions are derived and recognized for a Gibbs sampler, writing a loop to iteratively sampling each parameter is a helpful exercise and challenge. Similarly, once students have a good conceptual grasp of a Metropolis algorithm, implementing it with a few lines of code can be beneficial for developing their programming skills and deepening their understanding of MCMC.
After the basic tenets of MCMC algorithms are learned, students can use MCMC software for models requiring more advanced MCMC techniques. 

To illustrate different popular MCMC samplers, we introduce a three-parameter change point model of named storms in Section \ref{MCMC:example}. We then proceed to the Gibbs sampler in Section \ref{MCMC:Gibbs} and the Metropolis algorithm in Section \ref{MCMC:metropolis}, illustrated within the context of the change point example with sample R scripts. In Section \ref{MCMC:JAGS}, we present the use of JAGS, a popular MCMC software, within the context of the change point example with a sample JAGS script.

\subsection{A Change Point Example of Named Storms}
\label{MCMC:example}

\begin{figure}[H]
  \centering
\includegraphics[scale=0.3]{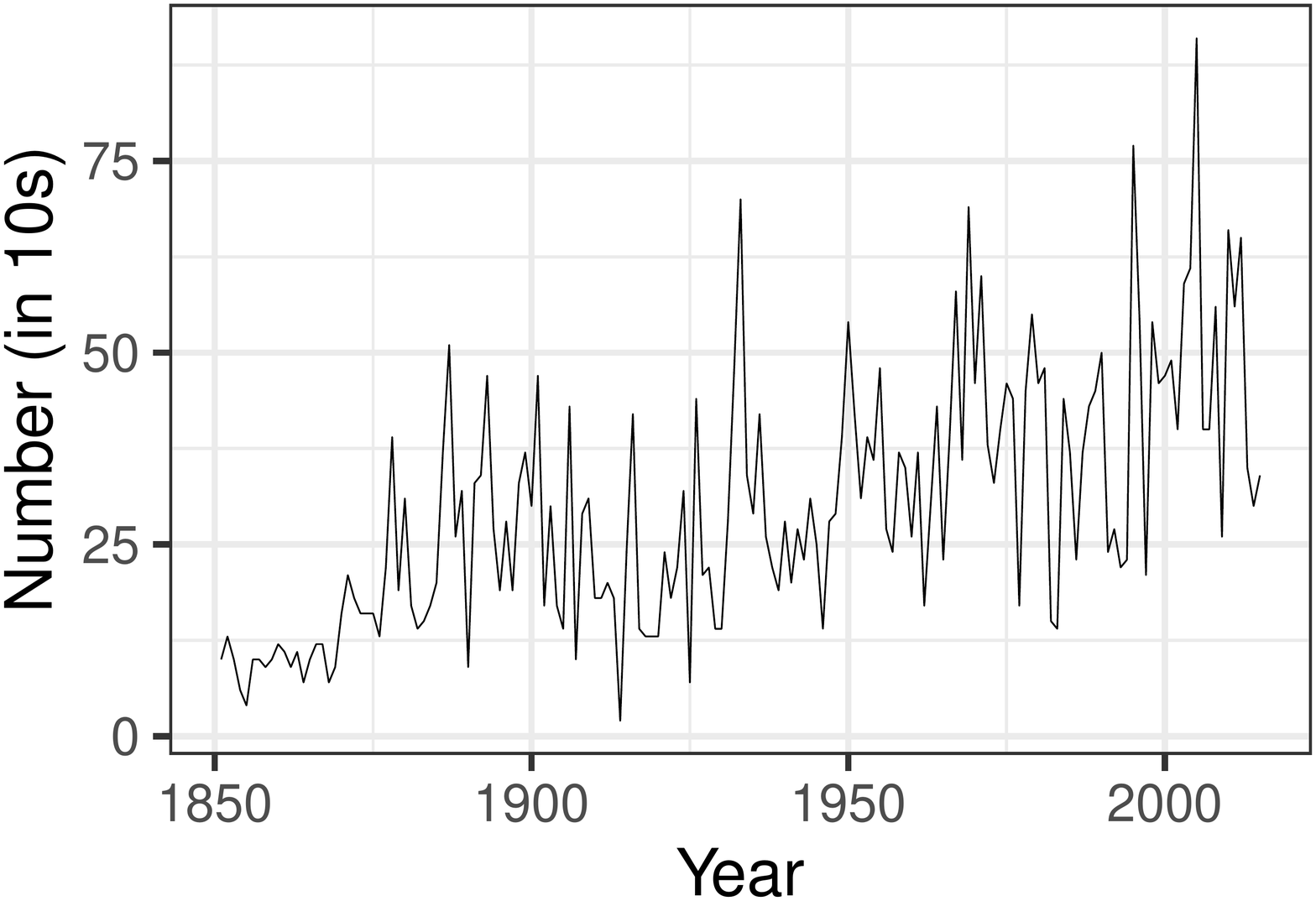}
\caption{\label{fig:exactapprox} The distribution of named storms over time, 1851 - 2015.}
\label{fig:StormsLine}
\end{figure}

The World Meteorological Organization maintains the practice of naming storms (tropical cyclones) so that they are quickly identified in warning messages. In a study on distribution of named storms over time, data has been collected on the number of named storms in Atlantic Ocean from 1851 to 2015, a period of 165 years.  Figure \ref{fig:StormsLine} displays the counts of named storms across years.  Suppose we are interested in whether the distribution of named storms has changed over time. If indeed there was a change in distribution, when did this change occur? 

Consider a model where we assume one and only one change point in time.  That is, assume there are two distinct periods: the  period before and the  period after the occurrence of the change point. Let $Y_i$ denote the number of named storms in year $i$, where $i = 1, \cdots, n = 165$ and let $M$ denote the year in which the distribution of random variable $Y$ changes, where $M \in \{1, \cdots, n - 1\}$.

Similar to our ED visits example in Section \ref{intro:examples}, we can use a Poisson sampling model for counts of named storms.  A Poisson sampling model has a single parameter $\lambda$, where $\lambda$  represents the average count of storms in that period. Our model assumes one and only one change point in time denoted by $M$. Conditional on $M$, the first period counts (before the change point occurred) follow a Poisson($\lambda_1$) sampling distribution, and the second period counts (after the change point occurred)  follow a Poisson($\lambda_2$) sampling distribution. Specifically,
\begin{eqnarray}
Y_i \mid \lambda_1, M &\overset{i.i.d.}{\sim}& \textrm{Poisson}(\lambda_1), \,\, i = 1, \cdots, M, \\
Y_i \mid \lambda_2, M &\overset{i.i.d.}{\sim}& \textrm{Poisson}(\lambda_2), \,\, i = M + 1, \cdots, n.
\end{eqnarray}

There are three parameters in our model: the two Poisson rates $\lambda_1$ and $\lambda_2$, and the change point $M$. We can then express the likelihood function by
\begin{eqnarray}
L(\lambda_1, \lambda_2, M) &=&  \prod_{i=1}^{M} f(y_i \mid \lambda_1, M) \prod_{i=M+1}^{n} f(y_i \mid \lambda_2, M) \nonumber \\
&=& \prod_{i=1}^{M} \frac{\lambda_1^{y_i} \exp(-\lambda_1)}{y_i!} \prod_{i=M+1}^{n} \frac{\lambda_2^{y_i} \exp(-\lambda_2)}{y_i!} \nonumber \\
%&\propto& \lambda_1^{\sum_{i=1}^{M}y_i} \exp\left(-\sum_{i=1}^{M}\lambda_1\right) \lambda_2^{\sum_{i=M+1}^{n}y_i} \exp\left(-\sum_{i=M+1}^{n}\lambda_2\right) \nonumber \\
&\propto& \lambda_1^{\sum_{i=1}^{M}y_i} \exp(-M \lambda_1) \lambda_2^{\sum_{i=M+1}^{n}y_i} \exp(-(n - M) \lambda_2).
\end{eqnarray}
With a  prior $g(\lambda_1, \lambda_2, M)$, we arrive at the joint posterior density:
\begin{eqnarray}
g(\lambda_1, \lambda_2, M \mid \{y_i\}) \propto \lambda_1^{\sum_{i=1}^{M}y_i} \exp(-M \lambda_1) \lambda_2^{\sum_{i=M+1}^{n}y_i} \exp(-(n - M) \lambda_2) \times g(\lambda_1, \lambda_2, M), \nonumber \\
\lambda_1 > 0, \lambda_2 > 0, M = 1, ..., n - 1. \nonumber
\end{eqnarray}

\subsection{The Gibbs sampler}
\label{MCMC:Gibbs}

A Gibbs sampler iteratively samples one parameter at a time given its full conditional distribution, defined as the parameter's distribution conditional on values of the remaining parameters. For all full conditional distributions to exist, we typically assign an independent conjugate prior for each parameter of interest. For $\lambda_1$ and $\lambda_2$, we can use conjugate gamma priors:
\begin{eqnarray}
\lambda_1 \mid \alpha_1, \beta_1 &\sim& \textrm{Gamma}(\alpha_1, \beta_1), \label{eq:lambda1_prior}\\
\lambda_2 \mid \alpha_2, \beta_2 &\sim& \textrm{Gamma}(\alpha_2, \beta_2). \label{eq:lambda2_prior}
\end{eqnarray}
For $M$, we can use a uniform prior over integers from 1 to $n - 1$ to represent little prior knowledge about the location of the change point $M$:
\begin{equation}
Pr(M = m) = \frac{1}{n - 1}, \,\, M \in \{1, \cdots, n - 1\}. \label{eq:M_prior}
\end{equation}

With these prior choices and some algebra, the joint posterior density of $\{\lambda_1, \lambda_2, M\}$ is given by
\begin{eqnarray}
g(\lambda_1, \lambda_2, M \mid \{y_i\}) &\propto& \lambda_1^{\sum_{i=1}^{M}y_i + \alpha_1 - 1} \exp(-(M + \beta_1)\lambda_1) \times \nonumber \\
&& \lambda_2^{\sum_{i=M+1}^{n}y_i + \alpha_2 - 1} \exp(-(n - M + \beta_2) \lambda_2).
\label{eq:ChangePointPost_Gibbs}
\end{eqnarray}

From Equation (\ref{eq:ChangePointPost_Gibbs}), we can recognize  the full conditional posterior distributions of $\lambda_1$ and $\lambda_2$, each of which is a gamma:
\begin{eqnarray}
\lambda_1 \mid \lambda_2, M, \{y_i\} &\sim& \textrm{Gamma}\left(\sum_{i=1}^{M}y_i + \alpha_1, M + \beta_1\right), \label{eq:lambda1_post}\\
\lambda_2 \mid \lambda_1, M, \{y_i\}&\sim& \textrm{Gamma}\left(\sum_{i=M+1}^{n}y_i + \alpha_2, n - M + \beta_2\right). \label{eq:lambda2_post}
\end{eqnarray}
Furthermore,  the full conditional posterior distribution of $M$ is  a discrete distribution over $m \in \{1, \cdots, n - 1\}$:
\begin{equation}
Pr(M = m \mid \lambda_1, \lambda_2, \{y_i\}) \propto \lambda_1^{\sum_{i=1}^{m} y_i} \lambda_2^{\sum_{i=m+1}^{n}y_i} \exp((\lambda_2 - \lambda_1)m),
\label{eq:M_post}
\end{equation}
which can be sampled from a multinomial distribution with normalized probabilities over $m \in \{1, \cdots, n - 1\}$.

Below is a sample R script to run a Gibbs sampler for this example starting with an initial guess at the parameter $M$.  We use \texttt{rgamma()} and \texttt{rmultinom()} functions to sample from gamma and multinomial distributions respectively.   In the main loop structure, we first simulate $\lambda_1$ from its posterior in Equation (\ref{eq:lambda1_post}) and $\lambda_2$ from its posterior in Equation (\ref{eq:lambda2_post}).  We simulate $M$ from its posterior in Equation (\ref{eq:M_post}), where we work on the log scale and subtract a constant for computational stability.

\begin{verbatim}
for (i in 1:iter){
  ## draw lambda_1
  lambda_1 <- rgamma(1, sum(y[1:M]) + alpha_1, M + beta_1)
  ## draw lambda_2
  lambda_2 <- rgamma(1, sum(y[(M+1):n]) + alpha_2, n - M + beta_2)
  ## draw M
  term_m_log <- rep(NA, n - 1)
  subtract_term <- (log(lambda_1) + log(lambda_2)) * sum(y) / 2 +
    (lambda_2 - lambda_1) * n / 2
  for (m in 1:(n-1)){
    term_m_log[m] <- log(lambda_1) * sum(y[1:m]) + 
      log(lambda_2) * sum(y[(m+1):n]) + 
      (lambda_2 - lambda_1) * m - subtract_term
  }
  term_m <- exp(term_m_log)
  normalized_probs <- term_m / sum(term_m)
  M <- which(rmultinom(1, 1, normalized_probs) == 1)
}
\end{verbatim}

As shown in our illustration, to construct a Gibbs sampler, students need to write out the joint likelihood function and prior distributions of all parameters to derive and recognize each parameter's full conditional posterior distribution.  This derivation can be challenging for more complex models, but this process  deepens students' understanding of the Bayesian process of deriving the posterior from the likelihood and the prior. The implementation of Gibbs sampling for the change point example  should be somewhat familiar to students who have already seen Poisson sampling with a gamma prior. We acknowledge that the coding of the sampling of $M$ can be challenging. However, this Gibbs sampling exercise exposes students to not only new models but also new computing techniques, building up their skills that are transferrable to other settings.
In a similar fashion, we encourage illustrating the Gibbs sampler for  common inference problems such as the normal sampling model where both parameters are unknown,  normal sampling models with  hierarchical priors,  and  missing data and censored data problems such as presented in \citet{gelfand1990illustration}.  

Nevertheless, the derivation of a Gibbs sampler  does require a background in calculus and familiarity with the posterior derivations for popular sampling models with conjugate priors.  Since the Gibbs sampler is based on conditional probability distributions,  it would be challenging to communicate this algorithm to students with limited experience with conditional distributions.

\subsection{The Metropolis algorithm}
\label{MCMC:metropolis}

\subsubsection{Introduction}

A good general-purpose MCMC method is the Metropolis algorithm.  This algorithm can be viewed as a random walk through a posterior probability distribution. Without loss of generality, consider a single-parameter model with parameter $\theta$ and an observation vector $\bm y$.  Generically, let $\theta$ denote the parameter of interest with posterior density function $g(\theta \mid \bm{y})$.  If $\theta^c$ denotes the current simulated draw from the posterior, then  the Metropolis algorithm consists of three steps. First, we simulate a proposed value $\theta^p$ that is uniformly distributed from  $\theta^c - C$ to $\theta^c + C$, where $C$ is a positive constant specified by the user.  Next, we compute the ratio $R$ of the posterior densities at the proposed value $\theta^p$ and the current value $\theta^c$.  Last, we simulate a uniform random variate $U$: if the value of $U$ is smaller than $R$, then the algorithm moves to the proposed value, otherwise the algorithm will stay at the current value.  This random walk process defines a Markov chain and the collection of many iterations from this process will approximately be a sample from the posterior distribution. In addition to uniform, we can use a normal proposal distribution in step 1.

One advantage of the  Metropolis algorithm is that it is straightforward to program.  For example, the following snippet of R code implements \texttt{iter} number of iterations of the three steps of the algorithm for a posterior density of a single variable. In this code, 
\texttt{logpost()} is a function defining the log posterior function, \texttt{current} is the current value of $\theta$, \texttt{proposed} is the proposed value, $C$ is the half-width of the proposal interval, and \texttt{runif()} simulates a uniform random variate between two values.
\begin{verbatim}
for (i in 1:iter) {
    proposed <- runif(1, min = current - C,  max = current + C)
    R <- exp(logpost(proposed) - logpost(current))
    accept <- ifelse(runif(1) < R, "yes", "no")
    current <- ifelse(accept == "yes", proposed, current)
}
\end{verbatim}

\noindent {\underline{\bf Activity 4 - Random Walk on a Number Line}}: The ``Random Walk on a Number Line" activity in Section 1.4 of the Supplementary Material illustrates the  Metropolis algorithm in a simple setting.  Students define a probability mass function on the integers 1, 2, 3, 4, 5 and they implement several steps of the Metropolis algorithm using coin flips and a random number generator. We also provide an applet to visually show the output of the algorithm for a large number of iterations.

%It is helpful for students to write their own scripts to confirm and reinforce their understanding of the sampling procedure. In particular, they need to write their own \texttt{logpost()} function, defining the log posterior function of their model and prior. We now turn to our Metropolis implementation for the change point example of named storms.

\subsubsection{Metropolis within Gibbs Sampling}

The Metropolis algorithm can be used to simulate specific conditional distributions in Gibbs sampling when one or more full posterior conditional distributions do not have convenient functional forms for sampling.  Consider the use of the Gibbs sampling algorithm for the change point example, where we now use a non-conjugate normal prior for the Poisson rate parameter $\lambda_1$ while keeping the original gamma and uniform priors for $\lambda_2$ and $M$. That is, we assign the following prior for $\lambda_1$:
\begin{equation}
\lambda_1 \mid \mu, \sigma \sim \textrm{Normal}(\mu, \sigma). 
\end{equation}

Together with the gamma prior in Equation (\ref{eq:lambda2_prior}) for $\lambda_2$ and the uniform prior in Equation (\ref{eq:M_prior}), our joint posterior density of $(\lambda_1, \lambda_2, M)$ is given by
\begin{eqnarray}
g(\lambda_1, \lambda_2, M \mid \{y_i\}) &\propto&  \lambda_1^{\sum_{i=1}^{M}y_i} \exp(-M \lambda_1) \lambda_2^{\sum_{i=M+1}^{n}y_i} \exp(-(n - M) \lambda_2)\times  \nonumber \\
&& \frac{1}{\sigma \sqrt{2\pi}} \exp\left(-\frac{(\lambda_1 - \mu)^2}{2\sigma^2}\right) \lambda_2^{\alpha_2 - 1} \exp(-\beta_2 \lambda_2) \times \frac{1}{n - 1} \nonumber \\
&\propto& \lambda_1^{\sum_{i=1}^{M}y_i} \exp\left(-M\lambda_1 - \frac{(\lambda_1 - \mu)^2}{2\sigma^2}\right) \times \nonumber \\
&& \lambda_2^{\sum_{i=M+1}^{n}y_i + \alpha_2 - 1} \exp(-(n - M + \beta_2) \lambda_2).
\label{eq:ChangePointPost_Metropolis}
\end{eqnarray}

Following the work in Section \ref{MCMC:Gibbs},  the  gamma full posterior for $\lambda_2$ in Equation (\ref{eq:lambda2_post}) and the  discrete full posterior for $M$ in Equation (\ref{eq:M_post}) are obtained.  However, the full conditional posterior distribution of $\lambda_1$ does not have a familiar form:
\begin{equation}
g(\lambda_1 \mid \lambda_2, M, \{y_i\}) \propto \lambda_1^{\sum_{i=1}^{M}y_i} \exp\left(-M\lambda_1 - \frac{(\lambda_1 - \mu)^2}{2\sigma^2}\right).
\label{eq:lambda1_logpost}
\end{equation}
In such situations, the Metropolis algorithm can be used to sample the parameter $\lambda_1$. 

To code the Metropolis within Gibbs sampling algorithm, we need to write a function defining the log posterior function which is used for evaluating the ratio $R$ of the posterior densities at the proposed value $\lambda_1^p$ and the current value $\lambda_1^c$.  The sample script  function is based on the full conditional posterior distribution in Equation (\ref{eq:lambda1_logpost}).
\begin{verbatim}
logpost <- function(lambda_1){
  log(lambda_1) * sum(y[1:M]) - M * lambda_1 -
    (lambda_1 - mu) ^ 2 / (2 * sigma ^ 2)
}
\end{verbatim}

For this example, a  Metropolis step is coded for $\lambda_1$ while keeping the original Gibbs sampling code for the conditional distributions of  $\lambda_2$ and $M$.  For brevity,  we include only the snippet of R code for sampling $\lambda_1$ using a Metropolis step. The sampling of $\lambda_2$ and $M$ remains the same as in Section \ref{MCMC:Gibbs}. 

\begin{verbatim}
lambda_1_p <- runif(1, min = lambda_1_c - C,  max = lambda_1_c + C)
R <- exp(logpost(lambda_1_p) - logpost(lambda_1_c))
accept <- ifelse(runif(1) < R, "yes", "no")
lambda_1_c <- ifelse(accept == "yes", lambda_1_p, lambda_1_c)
\end{verbatim}

We ran the modified Gibbs sampler with a Metropolis step for our example.  We use a half-width value of $C = 2$ and the acceptance rate of the Metropolis algorithm for $\lambda_1$ is 0.382 which is within the recommended range of [0.2, 0.4] \citep{gelman1996efficient}. 

\noindent {\underline{\bf Exercises}}: After providing students with posterior derivation and the sample R script for implementing a Metropolis step, we can create follow-up exercises to help them practice. For example: (1) If we use a larger half-width value of $C$, how does it affect our acceptance rate? Verify your conjecture with $C = \{4, 8\}$; (2) Try a normal proposal distribution and compare results; (3) Can you think of another reasonable prior distribution for $\lambda_1$? If it is a non-conjugate prior, write code the associated \texttt{logpost()} function and the Metropolis step.

%For comparison, Figure \ref{fig:lambda1_trace} shows the traceplots of Gibbs and Metropolis of $\lambda_1$. Each MCMC sampler is run for 10,000 iterations with 5000 burn-in and 10 thin. We observe that after convergence, both algorithms provide similar posterior estimations of $\lambda_1$.

%\begin{figure}[H]
%  \centering
%\includegraphics[scale=0.3]{figures/GibbsMetropolis_lambda1.pdf}
%\caption{\label{fig:exactapprox} Traceplots of Gibbs and Metropolis outputs of $\lambda_1$.}
%\label{fig:lambda1_trace}
%\end{figure}

From a pedagogical perspective, conceptually the Metropolis algorithm is less challenging than  Gibbs sampling  as the algorithm consists of three intuitive steps:  (1) propose, (2) compute the acceptance probability, and (3) move or stay. In the process of programming this algorithm, students understand that it is helpful to work with the logarithm of the posterior density instead of the posterior density  for numerical stability. Moreover, students learn a helpful programming technique of adding a counter of acceptances to the program for monitoring the acceptance rate of the Metropolis algorithm. 
 
Moreover, the use of the Metropolis algorithm naturally leads to discussions about MCMC diagnostics.  
The performance of the algorithm depends on the width of the proposal region.   We can use specific MCMC diagnostic procedures such as traceplots and autocorrelation plots \citep{cowles1996markov} to illustrate good and poor  choices for the proposal region.  In addition, the Metropolis algorithm  is a generic algorithm as it works on a variety of Bayesian inference models.  Students'  learning and practice of the Metropolis algorithm not only enhances their statistical programming skills, but also deepens their overall understanding of MCMC.

On the negative side, the Metropolis algorithm requires tuning of the parameters in the proposal distribution, such as the width of the proposal region. Such tasks have to be done by trial and error.
For poor choices of the proposal region, the Metropolis algorithm can be slow in exploring the posterior parameter space. 
If students are not comfortable with programming, then coding the Metropolis algorithm may take class time that could be used instead to discuss statistical issues.  If there is a programming issue, the instructor could supply code and the focus would be on using this code for particular Bayesian modeling problems.

\subsection{Coding an MCMC Sampler Using JAGS}
\label{MCMC:JAGS}

Many software programs, such as JAGS \citep{plummer2003jags}, BUGS \citep{lunn2009bugs}, and Stan \citep{carpenter2017stan}, provide  MCMC samplers that can be applied for a large class of Bayesian modeling problems. The use of each program requires the writing of a Bayesian model script including the specification of the sampling model and the priors.  

For example, the following JAGS script can be used to specify our change point example. In this script, our three parameters are represented  in the script by the variables \texttt{lambda1}, \texttt{lambda2}, and \texttt{M}, respectively. The Poisson sampling is represented by the \texttt{dpois()} function, and the prior densities are specified by the \texttt{dgamma()} and \texttt{dunif()} functions, respectively (we use the same conjugate priors as in Section \ref{MCMC:Gibbs}). Note the use of \texttt{ifelse()} function in the sampling portion of the model, where we use a new parameter \texttt{lambda[i]} to store the selected rate parameter given the value of \texttt{M}.

\begin{verbatim}
modelString <-"
model {
for (i in 1:n){
lambda[i] = ifelse(i < M, lambda1, lambda2)
y[i] ~ dpois(lambda[i])
}
lambda1 ~ dgamma(alpha1, beta1)
lambda2 ~ dgamma(alpha2, beta2)
M ~ dunif(1, n - 1)
}"
\end{verbatim}

The \texttt{runjags} package provides an R interface to JAGS \citep{runjags}. Suppose the model script is contained in the variable \texttt{modelString} and the data is contained in the data frame \texttt{the\_data}.  Then the function \texttt{run.jags()}  implements the MCMC sampling as follows:

\begin{verbatim}
posterior <- run.jags(modelString,t
                      n.chains = 1,
                      data = the_data,
                      monitor = c("lambda1", "lambda2", "M"),
                      adapt = 1000,
                      burnin = 5000,
                      sample = 500,
                      thin = 10)
\end{verbatim}
In the function, we indicate by \texttt{n.chains = 1} that one stream of simulated draws will be run and  the \texttt{monitor = c("lambda1", "lambda2", "M")} argument indicates that we wish to collect simulated draws of $\{\lambda_1, \lambda_2, M\}$.  The remaining arguments say that 1000 iterations will be used in the adaptive phase to choose the appropriate MCMC algorithm, 5000 iterations will be used in a burn-in phase, and 500 samples will be collected in the sampling phase after thinning by every 10th draws.  Part of the output of the \texttt{posterior} object is a 500-by-3 matrix of simulated draws of $\{\lambda_1, \lambda_2, M\}$ that can be summarized for posterior inferences.  The posterior inferences from the simulated draws this use of JAGS are very similar to the inferences from the draws from the Gibbs sampling algorithm. Moreover, MCMC diagnostics plots can be easily generated from JAGS output with the \texttt{plot(posterior, vars = "")} command. 

\noindent {\underline{\bf Exercises}}: After providing students with the sample JAGS script, we can create follow-up exercises to help them practice writing JAGS scripts. For example: (1) Run 2 chains and compare the two chains; (2) Use a normal prior for $\lambda_1$ as in Section \ref{MCMC:metropolis} and compare results. Follow-up exercises on posterior inference will deepen students' understanding and hone their applied skills. For example, how do the posterior distributions of $\lambda_1$ and $\lambda_2$ different from each other, and what conclusions can we draw from these differences?

%\begin{figure}[H]
%  \centering
%\includegraphics[scale=0.3]{figures/GibbsMetropolisJAGS_lambda1.pdf}
%\caption{\label{fig:exactapprox} Traceplots of Gibbs, Metropolis, and JAGS outputs of $\lambda_1$.}
%\label{fig:lambda1_all_trace}
%\end{figure}

From a pedagogical perspective, the use of scripting languages such as JAGS allows students to implement posterior inference and MCMC samplers for more sophisticated Bayesian models such as multilevel models.
Texts such as {\it Bayesian Statistical Methods} \citep{BayesianModeling2019book} and {\it Probability and Bayesian Modeling} \citep{ProbBayes2019book} provide JAGS scripts for both basic and more advanced Bayesian models. 
In non-statistics fields, {\it Doing Bayesian Analysis} \citep{DoingBayesianAnalysis2015book} illustrate the use of JAGS and Stan scripts with a primary focus on students in psychology, cognitive science, social sciences, clinical sciences, and consumer sciences in business.

\section{Concluding Remarks}
\label{conclusion}
This paper has surveyed a wide range of computational methods available for instructors who wish to teach Bayesian methods at the undergraduate  level.  Our intent is to provide guidance on the computational methods to use that will help achieve the learning outcomes  of a Bayesian course as stated in Section \ref{intro:LOs}.  Non-simulation methods are illustrated for several Bayesian one- and two-parameter problems, while various MCMC samplers are illustrated with a three-parameter problem, all of which have been used in our teaching. A supplemental R Markdown file is available that will reproduce all of the Bayesian calculations described in this paper.
A second supplemental file illustrates the use of these computational methods within the context of specific activities with interesting data.  Students in our classes have found these activities helpful for understanding the computational algorithms and applying these algorithms in making Bayesian inferential and predictive conclusions.

For students with modest mathematical backgrounds, the use of discrete prior distributions is a helpful non-simulation approach for introducing the Bayesian paradigm in problems with one parameter. Conjugate analyses are also helpful in communicating Bayesian thinking when applicable.  Conjugate priors facilitate students' learning and practice of prior assessment and how the prior and data are combined in a posterior analysis. 

Estimation of multi-parameter Bayesian models is challenging, which motivates the use of MCMC simulation algorithms for sampling from the posterior distribution. For teaching the fundamental principles of MCMC, we advocate  introducing the Gibbs sampler to students with calculus backgrounds and an understanding of conjugate priors.  The  derivation of a Gibbs sampler with the identification of the relevant conditional distributions is a useful pedagogical exercise for these students. For introductory level courses for undergraduate students, we advocate introducing the Metropolis algorithm, as this MCMC sampler is easy to understand and naturally leads to a discussion about  the MCMC diagnostics methods to detect convergence of the algorithm.

For fitting advanced Bayesian models, we believe the use of MCMC software programs  such as JAGS reinforces students' learning.  JAGS is not quite a computational black box since conjugate posteriors, Gibbs sampling, and Metropolis sampling are the main MCMC methods incorporated into JAGS and these methods can be introduced to students.  In addition,  writing a model script in a software program such as JAGS requires a clear understanding of the sampling model and the prior. 

Generally students in our classes have found it helpful to have a straightforward and slow integration of Bayesian computing through discrete Bayes and conjugate analyses,  transitioning to simulation-based methods by comparison to exact conjugate analyses solutions. With regards to MCMC, the students benefit with contrasting different MCMC samplers, and transitioning to MCMC software such as JAGS for advanced Bayesian models. Students' performances in computing labs and projects have demonstrated the effectiveness of our reviewed and proposed design of Bayesian computing techniques in the undergraduate curriculum.

Fortunately, there are a number of R packages such as \texttt{LearnBayes} (\citet{learnbayes}) and \texttt{runjags} that facilitate the use of these computational methods in Bayesian education at the undergraduate level.  The decision by the instructor on the methods to use will depend on the mathematical backgrounds of the students, the types of statistical problems considered, and the particular learning objectives of the Bayesian course or module.

%The use of wrapper functions such as \texttt{stan\_glm()} implement ``black-box" MCMC sampling methods which are potentially attractive to introductory-level undergraduate courses and applied Bayesian courses in non-statistics fields.

%Some ideas:
%
%\begin{itemize}
%
%\item discrete Bayes helpful for introducing the Bayesian paradigm
%\item conjugate priors helpful for prior assessment and understanding how the prior and data are combined in a posterior analysis
%\item derivation of Gibbs sampling is a useful pedagogical exercise
%\item scripts are helpful for going from model specification to code
%\end{itemize}

\section*{Acknowledgements}

We are very grateful to the editor, the associate editor, and four reviewers for their useful comments and suggestions.

\bibliography{BayesCompbib}

\end{document}

\def\spacingset#1{\renewcommand{\baselinestretch}%
{#1}\small\normalsize} \spacingset{1}

\if0\blind
{
\title{Supplementary Material for Bayesian Computing in the Undergraduate Statistics Curriculum}
\author{Jim Albert* and Jingchen Hu**\\
*Department of Mathematics and Statistics, Bowling Green State University\\
**Department of Mathematics and Statistics, Vassar College}

\maketitle
} \fi

\if1\blind
{
  \bigskip
  \bigskip
  \bigskip
  \begin{center}
    {\LARGE\bf Bayesian Computing in the Statistics and Data Science Curriculum}
\end{center}
  \medskip
} \fi

\bigskip

\begin{abstract}

This Supplementary Material contains four learning activities introduced in the main text, and R code for Section 2.3 Normal Approximation.
\end{abstract}

\section{Learning Activities}
\label{LA}

\subsection{Discrete Bayes Activity: ``Learning About a Proportion"}
\label{LA:discrete}

\subsubsection*{Introduction}

In this activity, you will gain experience in constructing a discrete prior distribution for a proportion that reflects your beliefs about the location of the proportion.

\subsubsection*{Helpful Apps}

The ``Learning About a Proportion Using Bayes' Rule" app at \newline
\url{https://bayesball.github.io/nsf\_web/jscript/p\_discrete/prior2a.htm}

\noindent is helpful in computing posterior probabilities using a discrete prior.

\subsubsection*{Story}

Suppose you are interested in the proportion $p$ of students from your campus who need corrective vision.  {\bf Note:} This activity can be adjusted to learn about any proportion that might be of interest to your students.

\subsubsection*{Part 1:  Choosing a Prior}

\begin{enumerate}
\item Make a short list of plausible values for $p$.  In the event that you can't easily construct this list, use the eleven values $p$ = 0, 0.1, 0.2, ..., 0.9, 1.  Write down the following table with columns $p$, Weights and Prior and place your values of $p$ in the $p$ column.

\begin{center}
\begin{tabular}{|c|c|c|} \hline
$p$ &Weight & Prior \\ \hline
& & \\ \hline
& & \\ \hline
& & \\ \hline
& & \\ \hline
& & \\ \hline
\end{tabular}
\end{center}

\item Assign a weight of 10 to the value of the proportion $p$ that is most likely.  Put this weight value in the table.

\item Assign weights of 5 to those values of the proportion $p$ that you believe are half as likely as the value of $p$ that you selected in part 2.  In a similar manner, assign integer weight values (say 1, 2, 3, 4, 5, 6, 7, 8, or 9) to the other values of $p$. Place all of these weight values in your table.

\item Compute the sum of the weight values and put this sum at the bottom of the Weight column.

\item Find the probabilities for your prior by dividing each weight values by the sum of the weights.

\item Using your prior, find the probability that the proportion of students needing corrective vision is over 0.5.

\item Find the probability that the proportion of students needing corrective vision is at most 0.3.

\end{enumerate}

\noindent {\bf Part 2:  The Posterior}

\begin{enumerate}
\item Suppose you collect data from a sample of 20 students.  Of this sample, 13 students need some type of corrective vision.  Using Bayes' rule find the posterior of the proportion $p$.

\item Display your prior and posterior probabilities on the same scale.  Describe how your prior opinions have changed in the light of this new information.

\item Using your posterior, compute the probability that the proportion is over 0.5, and the probability that the proportion is at most 0.3.  Describe how these probability computations have changed from the prior to the posterior.
\end{enumerate}

\subsection{Conjugate Prior Activity ``Did Shakespeare Use Long Words?"}
\label{LA:conjugate}

\subsubsection*{Helpful Apps}

The "Constructing a Beta(a, b) Prior From Two Quantiles" at 

\url{https://bayesball.shinyapps.io/ChooseBetaPrior\_3/ }

\noindent is helpful in constructing a conjugate beta prior for a proportion.

\subsubsection*{Introduction}

One way to measure the complexity of some written text is to look at the frequency of long words, where we will define a "long word" as one that has 7 or more characters. Actually, we are interested in the fraction of all words that are long. For example, consider this sentence (from Moby Dick):

\medskip

\noindent ``These {\bf reflections} just here are {\bf occasioned} by the {\bf circumstance} that after we were all seated at the table, and I was {\bf preparing} to hear some good {\bf stories} about {\bf whaling}; to my no small {\bf surprise}, nearly every man {\bf maintained} a {\bf profound} {\bf silence}."

\medskip

\noindent There are a total of 41 words of which 10 (the ones in bold type) are long, so the fraction of long words is 10/41 = 0.24.

Let $P$ denote the proportion of long words among all of the plays written by William Shakespeare.

\subsubsection*{Part 1: Choosing a Prior}

\begin{enumerate}
\item Without looking at any Shakespeare text, make an educated guess at the value of $P$.  You will be specifying the median, the value $M$ such that it is equally likely that $P$ is smaller or larger than $M$ (that is, $Prob(P < M) = 0.5$) .

\item Without looking at any Shakespeare text, find the 90th percentile $P_{90}$ such that your prior probability that $P$ is smaller than $P_{90}$ is equal to 0.90 (that is, $Prob(P < P_{90}) = 0.90$).

\item Based on your answers to questions 1 and 2, use the app at 

\url{https://bayesball.shinyapps.io/ChooseBetaPrior\_3/ }

\noindent to find the shape parameters of your beta prior that match your statements about the values of $M$ and $P_{90}$.

\item Using the app, find the values of $P$ that bracket the middle 50\% of the prior probability, and the values of $P$ that bracket the middle 90\% of the prior probability.  Put these values below:

50\% interval: \_\_\_\_\_\_\_\_\_\_\_\_

90\% interval:  \_\_\_\_\_\_\_\_\_\_\_\_

\item Reflecting on the 50\% and 90\% intervals, are you interested in changing your statements about the values of $M$ and $P_{90}$?  If so, adjust your values of $M$ and $P_{90}$ and find your new values of the shape parameters of your beta prior.

\end{enumerate}

\noindent {\bf Part 2:  The Posterior Analysis}

\begin{enumerate}

\item Now collect some data. Going to

\texttt{http://shakespeare.mit.edu/}

choose one play and select approximately 100 words from your chosen play.  Paste your selection of words to the site 

\texttt{https://wordcounttools.com/ }

This site will count the number of words in your text and also give you the count of long words. Record the number of words $N$ and number of long words $Y$ you find.

\item Find the shape parameters of the beta posterior for $P$ that combines your prior found above with the data information.

Using the  
\texttt{qbeta()} and \texttt{pbeta()} functions in R or the app at

 \url{https://homepage.divms.uiowa.edu/~mbognar/applets/beta.html} 
 
\noindent to answer questions 8 through 10.

\item Find the posterior median.

\item Find the posterior probability that $P$ is larger than 0.20.

\item  Find a 90 percent interval estimate for $P$.

\end{enumerate}

\subsection{Activity:  Normal Approximation to Posterior}
\label{LA:normal}

\subsubsection*{Introduction}

This activity explores the accuracy of the normal approximation in a situation where the actual posterior density is not normal in shape.

\subsubsection*{Description}

Suppose you are interested in learning about the standard deviation in the time that it takes to commute to work.  You collect the the following times (in minutes) for 8 trips:

11  2 10  7  8  5  9  5  6  9

Assume that your commuting time $y$ is normally distributed with known mean of 10 seconds and standard deviation $\sigma$.  Assuming that your commuting times are independent, the likelihood function of $\sigma$ is equal to
$$
L(\sigma) = \prod_{j = 1}^{8}\frac{1}{\sigma} 
\exp\left(-\frac{1}{2 \sigma^2}(y_j - 10)^2\right)
$$
If you place a uniform prior on $\sigma$, then the posterior density of $\sigma$ is proportional to:

$$
g(\sigma | y) = \frac{1}{\sigma^8}\exp\left(-\frac{1}{2 \sigma^2}\sum_{j=1}^8(y_j - 10)^2\right), \, \, \sigma > 0
$$

For these data, $\sum_{j=1}^8(y_j - 10)^2 = 146$, so the posterior density is proportional to:

$$
g(\sigma | y) = \frac{1}{\sigma^8}\exp\left(-\frac{1}{2 \sigma^2}146\right), \, \, \sigma > 0
$$

\begin{enumerate}
\item Graph this posterior density over the interval (0, 14).  Describe the shape of this curve.  Would it be appropriate to approximate this density with a normal curve?

\item Compute the posterior density on the grid of values of 0.1, 0.2, ..., 13.9, 14.  Using these values, compute the posterior mean and posterior standard deviation.  These will be good approximations to the actual posterior mean and posterior standard deviation.

\item Find a normal approximation to this posterior density.  One way is to write a short function in R defining the logarithm of the posterior and using a function such as ```laplace()``` in the ProbBayes package to find the mean and standard deviation of the normal approximation.

\item Redraw the exact posterior density from part 1 and overlay the normal approximation curve.  Comment on the accuracy of the normal approximation.

\item Compare the "exact" posterior mean and posterior standard deviation with the values found from the normal approximation.

\item Suppose you are interested in the posterior probability $P(\sigma > 8)$.  Compute this probability two ways, one using the grid of values from part 2 and one using the normal approximation.  Comment on the accuracy of the normal approximation.

\item Suppose you are interested in computing a 90\% interval estimate of $\sigma$.  Compute this interval two ways, one using the grid of values from part 2 and one using the normal approximation.  Comment on the accuracy of the normal approximation.

\end{enumerate}

\subsection{Metropolis Activity ``Random Walk on a Number Line"}
\label{LA:metropolis}

\subsubsection*{Introduction}

This activity illustrates the Metropolis algorithm for sampling from a probability mass function defined on a number line.

\subsubsection*{Helpful Apps}

The ``Metropolis Random Walk" app at 
\url{https://bayesball.shinyapps.io/Metropolis/ }
is used to show a Metropolis random walk for a probability distribution on a small set of integer values.

\subsubsection*{Description}

Suppose we define the following probability mass function on the values 1, 2, 3, 4, 5.  A graph of this probability mass function is displayed.

\begin{figure}[H]
  \centering
\includegraphics[scale=0.3]{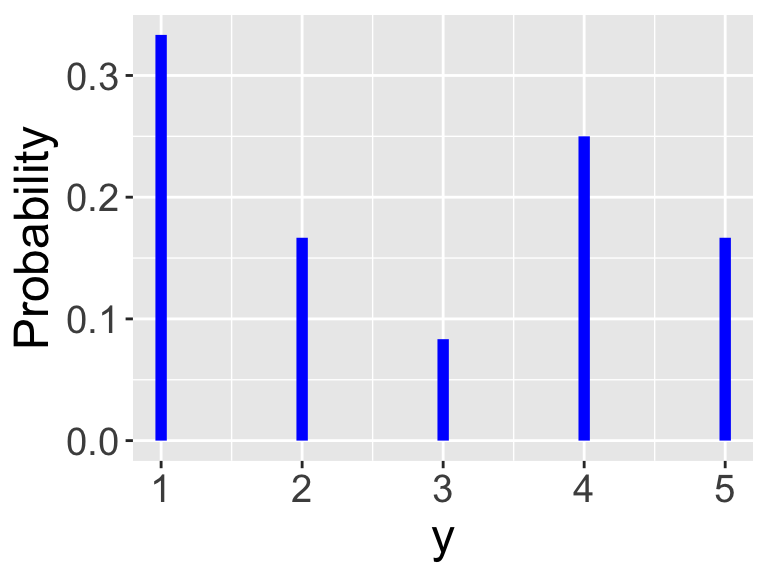}
\caption{A probability mass function defined on five discrete values.}
\end{figure}

Suppose we sample from this distribution in the following way.

\begin{itemize}
\item Step 1.  Start at any possible value of $Y$, say $Y = 2$.  This is the Current location.

\item Step 2.  Flip a coin -- if it lands heads, then the Candidate is the value one step to the left; if the coin lands tails the Candidate is the value one step to the right.

\item Step 3.  Compute the ratio 
$$
R = \frac{weight(Candidate)}{weight(Current)}.
$$

(Note for ease of calculation, I am using the weights in the table rather than the actual probabilities to compute this ratio.)

\item Step 4.  Generate a random number $U$ from 0 to 1.  If the value of $U$ is less than the ratio $R$, move to the Candidate location.  Otherwise, remain at the Current location.
\end{itemize}

Repeat steps 2 through 4 many times.  After many iterations, the relative frequencies of the visits to the five different values of $y$ will approximate the probability distribution.

\begin{enumerate}

\item  If you have a coin and a method for simulating a random number from 0 to 1, you can physically implement this random walk simulation.  Starting at location $y = 2$, take 10 steps of this simulation, recording all of your visits.

\item A Shiny app for illustrating this random walk simulation can be found at 
\url{https://bayesball.shinyapps.io/Metropolis/}

Using this app ...

\begin{enumerate}
\item Set the weights of your probability distribution to 4, 2, 1, 3, 2.

\item Set the number of iterations of the simulation to 100.

\item This app will display a graph of the simulated locations as a function of the iteration number.  Looking at this graph, what are some interesting features?

\item The bottom graph in this app shows a histogram of the simulated locations.  Set the number of iterations to 500.  Is the shape of this histogram similar to the shape of the probability density?   Is this what you would expect?  Explain.
\end{enumerate}
\end{enumerate}

\section{R Code for Section 2.3 Normal Approximation}
\label{Rcode}
The calculations for the two-group logistic model example are facilitated by use of the  \texttt{LearnBayes} package \citep{albert2018package}.  First we write a short function \texttt{logistic\_posterior()} that computes the logarithm of the posterior density of $\bm \beta = (\beta_0, \beta_1)$ in Equation (\ref{eq3}).
\begin{verbatim}
logistic_posterior <- function(theta, df){
  beta0 <- theta[1]
  beta1 <- theta[2]
  lp <- beta0 + beta1 * df$female
  p <- exp(lp) / (1 + exp(lp))
  sum(df$s * log(p) + df$f * log(1 - p)) +
    dcauchy(beta1, 0, 0.5, log = TRUE) +
    dnorm(beta0, 0, sqrt(1 / 0.0001), log = TRUE)
}
\end{verbatim}

Suppose we observe $y_M = 8$ Facebook users in a sample of $n_M = 30$ men, and $y_W = 15$ Facebook users in a sample of $n_W = 30$ women.  A data frame \texttt{ldata}  is constructed that contains the numbers of successes (i.e. users), the numbers of failures (i.e. non-users), and the female indicator variables for the two groups.  We find the normal approximation by use of the \texttt{laplace()} function\footnote{The \texttt{laplace()} uses the general-purpose \texttt{optim()} function from the \texttt{stats} package in R.}:
\begin{verbatim}
fit <- laplace(logistic_posterior, c(0, 0), ldata)
\end{verbatim}

The object \texttt{fit} from \texttt{laplace()} function has several outputs.  The vector \texttt{fit\$mode} gives the posterior mode of the posterior and \texttt{fit\$var} provides the estimate at the posterior variance-covariance matrix of the parameter vector.

For most of the Bayesian models considered at the undergraduate level including regression, the Laplace approximation is applicable and provides reasonable suitable approximations to the posterior distribution.

\bibliography{BayesCompbib}